\documentclass[runningheads]{svmult}

\usepackage{makeidx}   
\usepackage{graphicx}  
\usepackage{subeqnar}  
\usepackage{multicol}  
\usepackage{physprbb}  
\makeindex             

\def\ltsima{$\; \buildrel < \over \sim \;$}
\def\simlt{\lower.5ex\hbox{\ltsima}}
\def\gtsima{$\; \buildrel > \over \sim \;$}
\def\simgt{\lower.5ex\hbox{\gtsima}}

\begin{document}
\title*{The Cosmic Star-Formation History:\protect\newline 
The UV finds most}
%
%
%
%
\titlerunning{Cosmic SF History: UV}
%
\author{Kurt Adelberger}
%
%
%
\institute{Harvard-Smithsonian Center for Astrophysics, 60 Garden St., Cambridge MA 02138, USA}

\maketitle              


\section{Introduction}
The assigned subtitle of this talk was ``The UV Takes it All!''
The absurdity of that subtitle must have been obvious to many conference
participants when they first read the program, and now, after
four days of talks, I hope it is obvious to everyone here.
The UV cannot possibly ``take all'' of the cosmic star-formation history.  
Data from low and high redshift alike overwhelmingly show that
rapidly star-forming galaxies emit the bulk of their bolometric
luminosities in the far-infrared, not the UV.  Some extremely luminous high-redshift 
galaxies have not been detected in the UV at all (see, e.g.,
Hughes' and Blain's contributions to these proceedings); many others
dominate relatively short $850\mu$m exposures but
are barely detected in even the deepest (rest-frame) UV images.  I was asked
by the organizers to make a case for studying the distant universe
in the rest-frame UV, and I will; but I cannot pretend that UV surveys
are the best way to study all star-forming galaxies at high redshift.
Some fraction of the
stars in the universe formed in extremely dusty galaxies
that are best studied in the infrared.  Nevertheless the
available data---even IR data!---suggest that {\it most}
stars formed in objects that are easiest to detect in
the rest-frame ultraviolet, and so I have modified the subtitle
of this talk to a statement I am prepared to defend.
UV observations may not detect most of the luminosity emitted
by any single galaxy at high redshift, and may not detect
some high redshift galaxies at all, but the average high redshift
star-forming galaxy is much more easily detected through its
UV radiation than its dust emission.
In the first half of my talk I will lay out the arguments
supporting that statement.  In the second half I will
discuss the attempts to estimate star-formation
rates for UV-selected galaxies after they have been detected.
It is here (as we shall see) that dust obscuration poses
the greatest problems for UV surveys.

\section{The UV finds most}
\begin{figure}[b]
\begin{center}
\includegraphics[width=.55\textwidth]{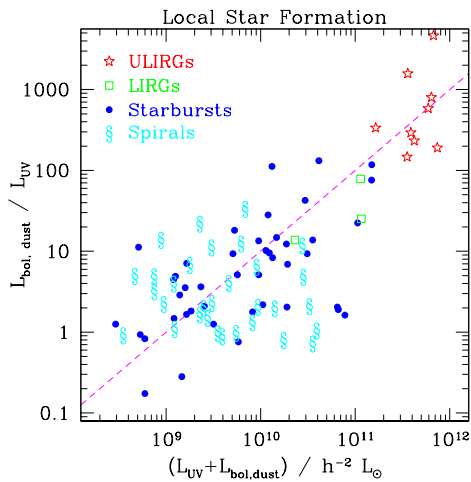}
\end{center}
\caption[]{Star formation in the local universe.  The quantity
on the abscissa is roughly proportional to star formation rate;
the quantity on the ordinate is a measure of dust obscuration.
Shown are ULIRGs from the samples of \cite{tks} and \cite{get},
starbursts from the sample of \cite{mhc}. and LIRGs and spirals
from the sample described in \cite{as}.  The trend of
decreasing obscuration with decreasing star-formation rate
is crudely fit by $L_{\rm bol,dust}/L_{\rm UV}\propto L_{\rm SFR}$
({\it dashed line\,})}
\label{localsf}
\end{figure}

Fig.~\ref{localsf} helps illustrate the argument I'm going to make.
The plot shows the level of dust obscuration observed
among different types of star-forming galaxies in the local universe.
The abscissa, $L_{\rm bol,dust}+L_{\rm UV}\equiv L_{\rm SFR}$, is the sum
of the star-formation luminosity that is and that is not
absorbed by dust; this provides a rough measure of total
star-formation rate.  The ordinate,
$L_{\rm bol,dust}/L_{\rm UV}$, is the ratio of obscured to
unobscured star-formation luminosity; it provides
a rough measure of the level of dust obscuration.
Here $L_{\rm bol,dust}\simeq 1.5 L_{\rm FIR}$ is
an estimate of a galaxy's bolometric dust luminosity
and $L_{\rm UV}\equiv \lambda l_\lambda$ evaluated at
$\lambda\sim 1600$\AA.  The well known trend of increasing dust obscuration with
increasing star-formation rate is obvious.  The expression
$L_{\rm bol,dust}/L_{\rm UV} \propto L_{\rm SFR}$ (dashed line) provides
a crude fit to the data.  Some galaxies in
the local universe are very heavily obscured ($L_{\rm bol,dust}/L_{\rm UV}>100$),
but these galaxies are rare and together they host only $\sim 5$\% of
the known local star-formation\cite{sm}.  The vast majority
of local star-formation occurs among less luminous and
less obscured spiral and starburst galaxies.

Now imagine taking the local universe and placing it at redshift
$z\sim 3$ (say) and observing it at $850\mu$m with SCUBA.
The only galaxies bright enough to detect would be the 
heavily dust-obscured
ultraluminous infrared galaxies (ULIRGs) with 
$100<L_{\rm bol,dust}/L_{\rm UV}<5000$.  When analyzing
this hypothetical $850\mu$m data it would be natural to
conclude that all high-redshift star formation occurred
in galaxies similar to the ones that were
detected, in extremely dusty galaxies that would be
difficult to detect in the rest-frame UV.
But this conclusion would be wrong.  It relies on the false
assumption that the comparatively faint galaxies which
host most of the star formation would be as dusty
as the ultraluminous galaxies that were detected;
it neglects the strong correlation of dust opacity and star-formation rate.

The real situation has obvious similarities to 
this hypothetical case.  
The typical galaxy detected by SCUBA appears to be
heavily obscured by dust, orders
of magnitude brighter in the (rest-frame) far-IR than the UV.
But only $\sim$25\% of the $850\mu$m background is produced
by objects brighter than SCUBA's 2mJy 
detection limit\cite{bcs}, and only $\sim$8\% by objects
bright enough to be included in the $\sim 5$mJy samples
of \cite{bcr} and \cite{crlwb}---the only SCUBA samples
with well established optical counterparts\footnote{see also \cite{siobk}
and \cite{iet}, which
discuss sub-mm samples selected in a different way but
with the similarly bright typical $850\mu$m fluxes.}.
The 5mJy sub-mm/radio samples have provided the strongest support
for the popular belief that most high-redshift star formation
occurred in extremely dusty galaxies that are not
detected in UV surveys, but
the fraction of high-redshift star formation that occurs
in the extremely dusty objects of the 5mJy sub-mm/radio
samples is as small as the fraction of local star formation
that occurs in LIRGs and ULIRGs.  Should we believe
that objects in the 5mJy samples are any more representative
of typical star-forming galaxies at high redshift
than LIRGs and ULIRGs are of star-forming galaxies
in the local universe?

The data suggest that they are not.  In the remainder of
this section I will argue that dust opacity and
star-formation rate are as strongly correlated at
high redshift as in the local universe, and that this implies
that the fainter galaxies responsible for most of the
$850\mu$m background are much less obscured in the
UV than typical objects detected by SCUBA, sufficiently
unobscured, in fact, that their UV emission is easier to
detect than their dust emission.

The correlation of dust opacity and star-formation rate can
be established at $z\sim 1$ by comparing galaxies
selected at $850\mu$m with SCUBA to those selected at
$15\mu$m with ISO.  Owing to the strength of the
7.7 and $8.6\mu$m PAH emission features, $z\sim 1$ star-forming
galaxies are comparatively bright at $15\mu$m and deep $15\mu$m
ISO images detect dusty $z\sim 1$ galaxies many times
fainter than those detectable with SCUBA.  
Fig. 2a compares the star-formation rates 
and dust opacities of five $850\mu$m sources
at $z\sim 1$ to those of the 17 ISO $15\mu$m sources
in the HDF with $0.8<z<1.2$.  The trend of decreasing
dust opacity with decreasing star-formation rate is clear;
the comparatively faint ISO sources are nowhere near as heavily
dust obscured as the brighter $\sim 5$mJy $850\mu$m sources.
Since neither $15\mu$m nor $850\mu$m
observations detect a very large fraction of $z\sim 1$
galaxies' bolometric dust luminosities, large multiplicative
corrections were required when estimating 
$L_{\rm bol,dust}$ for this plot
from the detected $15\mu$m or $850\mu$m fluxes.  I adopted
corrections that assume these $z\sim 1$ galaxies have dust
SEDs similar to those of rapidly star-forming
galaxies in the local universe.  Details can be found in \cite{as};
but the main result, the correlation of star-formation rate
and dust opacity, is insensitive to the adopted corrections:
altering $L_{\rm bol,dust}$ moves galaxies diagonally
on the plot, parallel to the correlation.

At higher redshifts, galaxies' PAH features are redshifted well outside
ISO's $15\mu$m window and another approach is required to
measure the dust opacity of galaxies significantly less luminous than the
5mJy SCUBA sources.  One approach, adopted by \cite{cet00}
and \cite{cet01}, is to point SCUBA at many galaxies which
are too faint to be detected individually, and sum
the measured fluxes.  The 33 Lyman-break galaxies at $z\sim 3$ targeted
by \cite{cet00} and \cite{cet01} were found to have
a mean $850\mu$m flux of $0.72\pm 0.31$mJy, corresponding
to $L_{\rm bol,dust}\sim 2.7\times 10^{11} h^{-2} L_\odot$
for the $\Omega_M=0.3$, $\Omega_\Lambda=0.7$ cosmology
adopted throughout.  The mean value of
$L_{\rm bol,dust}/L_{\rm UV}$ for the galaxies
in this sample was 13, significantly lower than
the values observed among brighter SCUBA sources at similar redshifts
(Fig 2b).
Since 0.7mJy is close to the typical flux of the sources
that dominate the $850\mu$m background\cite{bkis}\cite{bcs},
one might expect Lyman-break galaxies' moderate ratio of dust to UV luminosity
$L_{\rm bol,dust}/L_{\rm UV}\sim 10$ to be more representative
of average high-redshift star-forming galaxies
than the extreme ratios $L_{\rm bol,dust}/L_{\rm UV}\simgt 200$
of the bright $850\mu$m samples.  In any case, the $850\mu$m
observations of Lyman-break galaxies provide
further evidence that star-formation rate and dust opacity
are correlated at high redshift.

\begin{figure}[b]
\hbox to \hsize{ \centering
 \leavevmode
 \includegraphics[width=.47\textwidth]{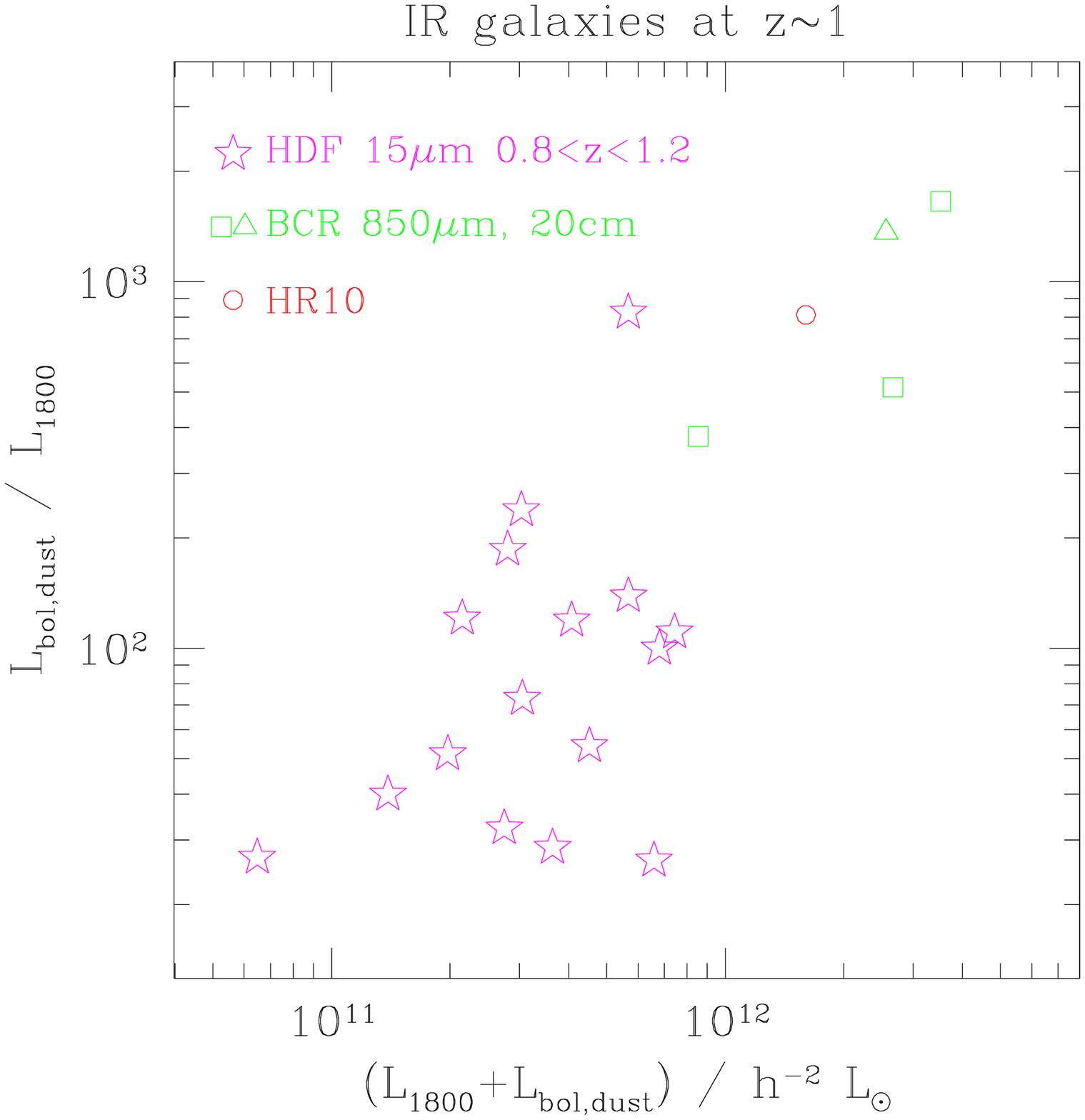}
 \hfil
 \includegraphics[width=.47\textwidth]{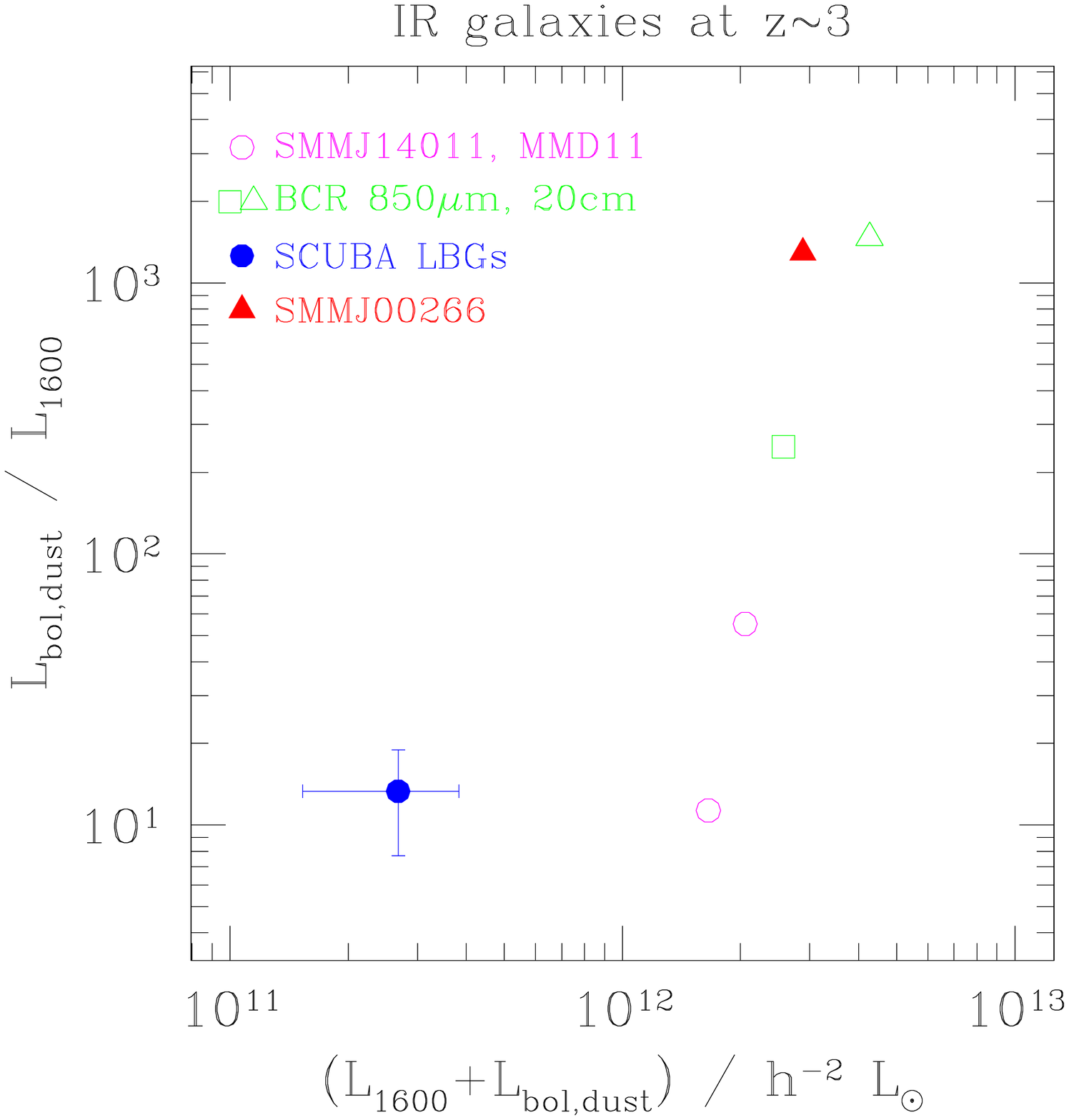}
}
\caption[]{The correlation of star-formation rate and dust opacity
at high redshift.  Left panel:  $z\sim 1$.  The bright
$850\mu$m sources from \cite{bcr} (BCR sub-sample with radio/sub-mm
photometric redshift $1<z_{\rm mm}<2$) and \cite{dey} (HR10) 
have larger ratios of dust to UV luminosity than the intrinsically fainter
$15\mu$m ISO sources from \cite{aces} (see also \cite{as}).
Right panel:  $z\sim 3$.  Bright $850\mu$m sources with
estimated/measured redshifts $z>2$ (from \cite{bcr},\cite{cet00},
\cite{fet99},\cite{fet00}) have larger ratios of dust to UV luminosity
than the (fainter) Lyman-break galaxies with similar redshifts
that have been statistically detected at $850\mu$m by \cite{cet01}
}
\label{highzsf}
\end{figure}

If we accept that dust opacity and star-formation rate
are correlated at high redshift in a way similar to
what Figs. 2a and b suggest, we can proceed to estimate the
characteristic level of dust obscuration among the
comparatively faint galaxies that dominate the
$850\mu$m background and see whether
most of them are easier to detect in the sub-mm or the UV.
Given a rest-frame UV detection limit of 
$f_\nu(1600$\AA$\times (1+z))>0.2\mu$Jy 
($m_{\rm AB}=25.5$),
roughly appropriate for the Lyman-break survey
of \cite{set99}, and a sub-mm flux limit of
$f_\nu(850\mu{\rm m})=2$mJy,
the confusion limit of SCUBA on the JCMT,
a galaxy will be easier to detect in the rest-frame UV
than the sub-mm if it has $L_{\rm bol,dust}/L_{\rm UV}\simlt 500$
at $z\sim 1$ or $L_{\rm bol,dust}/L_{\rm UV}\simlt 80$ at
$z\sim 3$\cite{as}.  Fig. 2a suggests that
$\langle L_{\rm bol,dust}/L_{\rm UV}\rangle=500$
is characteristic of galaxies with $L_{\rm SFR}\simeq 10^{12} h^{-2} L_\odot$
at $z\sim 1$.  Galaxies with lower star-formation luminosities
will be less dust obscured on average and easier to detect
in the UV than the sub-mm.  $10^{12} h^{-2} L_\odot$ corresponds
to $f_\nu(850\mu{\rm m})\simeq 3.3$mJy for $\Omega_M=0.3$, $\Omega_\Lambda=0.7$,
and the local dust SED shape assumed throughout; roughly 85\%
of the $850\mu$m background is produced by objects fainter
than 3.3mJy\cite{bcs}.  This suggests that the majority of star-formation
at $z\sim 1$ occurs in galaxies that are easier to detect
in the rest frame UV than the sub-mm.

A similar conclusion follows at $z\sim 3$.  Fig. 2b suggests that
$\langle L_{\rm bol,dust}/L_{\rm UV}\rangle=80$
is characteristic of galaxies with $L_{\rm SFR}\simeq 10^{12} h^{-2} L_\odot$
($f_\nu(850\mu{\rm m})\simeq 2.5$mJy) at $z\sim 3$.  80\% of
the $850\mu$m background is produced by objects fainter 
and presumably less obscured.

This completes my simple argument:
a correlation between star-formation rate and dust opacity
is seen to exist at every redshift we can probe ($0<z\simlt 3$),
and its slope is sufficient to imply that most of
the $850\mu$m background was produced by galaxies that
are easiest to detect in the rest-frame UV.
It is an empirical argument based
solely upon the observed ratios of dust to UV luminosity
among galaxies whose dust and UV emissions have both been
detected.  It largely sidesteps the long
and uncertain chain of reasoning -- the incompleteness corrections,
the luminosity-function extrapolations, the conversions of fluxes
to bolometric luminosities to star-formation rates, and
so on (see Carilli's contribution) -- that makes
attempts to address the same question through ``Madau diagrams''
so famously contradictory and unreliable.

\section{Interpreting UV surveys}
The reason UV surveys detect a large fraction of high redshift
star formation is that the UV luminosities of star-forming
galaxies are largely independent of dust opacity.
This is a consequence of the correlation of star-formation
rate and dust opacity:  the dustiest objects tend
to be the most luminous, and the two effects -- increased
obscuration and increased luminosity -- mostly cancel 
out at $\lambda\sim 1500$\AA.
As can be seen in Fig. 3, observed UV luminosities are
similar for galaxies with dust obscurations $L_{\rm bol,dust}/L_{\rm UV}$
spanning four orders of magnitude. 
ULIRGs in the local universe are as bright in the UV as 
starburst dwarfs; high redshift galaxies that have been
detected through their dust radiation are typically as bright in the
UV as those that have been detected through their UV radiation.

\begin{figure}[b]
\begin{center}
\includegraphics[width=.49\textwidth]{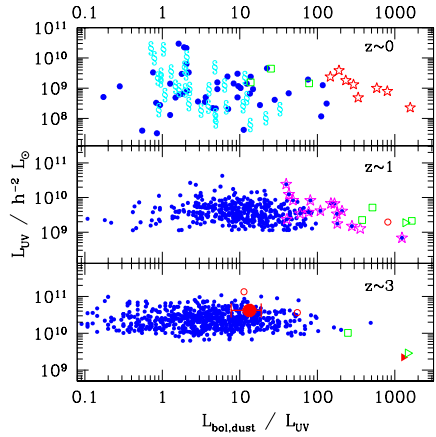}
\end{center}
\caption[]{Far-UV luminosity versus dust obscuration for star-forming galaxies
at $z\sim 0$ (top panel), $z\sim 1$ (middle panel), and $z\sim 3$ (bottom
panel).  The darker solid circles in the middle and bottom panel
represent UV-selected star-forming galaxies with $L_{\rm bol,dust}$
estimated from the $\beta$/far-IR correlation; the lighter solid
circle with error bars in the bottom panel shows the mean
UV luminosity and dust obscuration ($\pm$ standard deviation of
the mean) for Lyman-break galaxies observed in the sub-mm by
\cite{cet01}.  Otherwise the symbols are as in Figs. 1 and 2.  
In general extremely dusty galaxies
are no fainter in the far-UV than relatively dust-free galaxies
}
\label{fig17}
\end{figure}

This fact makes UV surveys of star-forming galaxies easy
to construct but hard to interpret.  Referring to
Fig. 3, one can see that a UV luminosity-limited survey of
the local universe would net a tremendously broad range
of objects, an indiscriminate haul of ULIRGs and blue compact dwarfs 
and spirals like the Milky Way.  The same appears to
be true of luminosity-limited surveys at high redshift.
But UV luminosities don't give us much to distinguish
between these vastly different objects:
if we were told only that a local galaxy had $L_{\rm UV}=10^9 h^{-2} L_\odot$,
for example, we would have no idea whether it was
a ULIRG or a spiral or a starburst dwarf.  How can
we tell what sort of objects we have detected in a high redshift UV
selected survey?
How can we estimate any single object's
star-formation rate to within even an order of magnitude?
Measuring mid-IR, far-IR, sub-mm, or radio fluxes for the
objects would be ideal, but most high-redshift star formation
occurs in objects too faint to be detected at any of these
longer wavelengths, and in any case the small fields of view
of current instruments at mid-IR to mm wavelengths ($\sim 2'\times 2'$) 
are poorly matched to the large fields of view
of modern optical instruments (up to $\sim 40'\times 40'$).
Until the next generation of long wavelength instruments becomes
available, much of our knowledge of galaxy formation at high
redshift will have to be derived from UV data alone.

Meurer and collaborators have developed one technique for estimating
dust luminosities (and hence star-formation rates) from UV observations
(e.g. \cite{mhc}, hereafter MHC).  
Dustier local starbursts are not fainter in the
UV on average, but they are redder, and MHC observed
that a galaxies' redness in the UV was a surprisingly
good predictor of its dustiness $L_{\rm bol,dust}/L_{\rm UV}$.  
Among the galaxies in
MHC's sample, a measurement of the UV spectral slope $\beta$
(in $l_\lambda\propto\lambda^\beta$
at $1200 \simlt \lambda \simlt 2000$\AA) was sufficient
to predict $L_{\rm bol,dust}/L_{\rm UV}$ to within
a factor of 2.  This $\beta$/far-IR correlation
is the basis of many attempts to estimate the
star-formation rates of UV-selected high-redshift galaxies.

Do high-redshift galaxies obey MHC's $\beta$/far-IR correlation?
The evidence is ambiguous.
In some cases the $\beta$/far-IR correlation appears to hold.
One example is
the lensed Lyman-break galaxy SMMJ14011+0252 at $z=2.565$.
Its observed $850\mu$m and 20cm fluxes of $15\pm 2$mJy, 
$115\pm 30\mu$Jy\cite{iet}
agree well with
the predictions of $23^{+14}_{-9}$ mJy,
$150^{+95}_{-67} \mu$Jy that the $\beta$/far-IR 
correlation implies\cite{as}.  In this case the total star-formation
rate can be deduced with reasonable accuracy from
UV observations alone.  The same appears to be
true for the ten brightest Lyman-break galaxies in
the HDF.  According to MHC, the $\beta$/far-IR and
far-IR/radio correlations predict a total 20cm
flux for these galaxies of $105\pm 24\mu$Jy, nicely consistent
with the total measured flux of $100\pm 33\mu$Jy.
But there are counter-examples as well.  Baker and
van den Werf discussed two in their talks, 
the galaxies MS1512+36-cB58 and MS1358+62-G1,
where the $\beta$/far-IR correlation
appears to overpredict the observed $850\mu$m flux by an order of
magnitude or more.  This suggests -- worryingly -- that failures
of the $\beta$/far-IR correlation at high redshift
are not restricted to the dustiest galaxies 
($L_{\rm bol,dust}/L_{\rm UV}\simgt 100$) as they appear
to be in the local universe (see Meurer's contribution
to these proceedings). 

Analysis of the 
largest available sample of UV-selected high-redshift galaxies
with independent constraints on $L_{\rm bol,dust}$
suggests that most high-redshift galaxies obey the
$\beta$/far-IR correlation better than  MS1512+36-cB58 and MS1358+62-G1
but perhaps not as well as SMMJ14011+0252.  Figure 4
shows the predicted and observed 20cm fluxes of
69 ``Balmer-break'' galaxies in the Hubble Deep and Flanking Fields with
spectroscopic redshifts $z=1.00\pm 0.11$\cite{as}.  The objects
that are predicted (via the $\beta$/far-IR and far-IR/radio correlations)
to have larger 20cm fluxes clearly do on average; the data
do not support the view that MHC's $\beta$/far-IR correlation
generally misestimates dust luminosities by more than
an order of magnitude (cf. van den
Werf's contribution to these proceedings).   
But nor do they obviously support the view that the $\beta$/far-IR
correlation can predict far-IR fluxes as accurately at high
redshift as in the local universe.  The measured 20cm fluxes
of the galaxies in the brightest bin of predicted flux
are roughly a factor of 3 times lower than expected, for example.
Assessing the significance of this shortfall
is difficult because of the large error bars in
the predicted fluxes -- a reasonable fraction of the
objects in the brightest predicted bin were probably scattered
out of the next brightest bin by various errors -- 
and in any case a significant discrepancy between the observed
and predicted fluxes might be due to a failure
of the far-IR/20cm rather than the $\beta$/far-IR
correlation.  But it is safe to say that the
available data do not inspire absolute confidence in the validity
of the $\beta$/far-IR correlation at high redshift. 

\begin{figure}[b]
\begin{center}
\includegraphics[width=.5\textwidth]{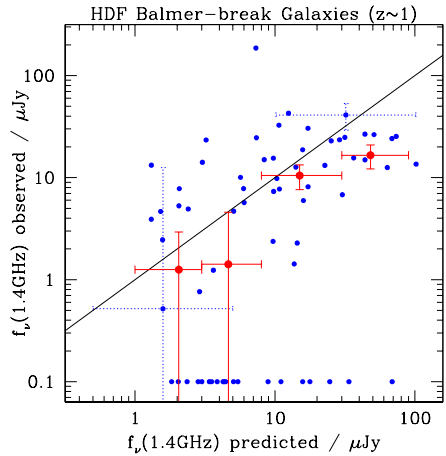}
\end{center}
\caption[]{Predicted and observed 20cm fluxes of $z\sim 1$ galaxies
in the HDF.  Small solid circles represent individual galaxies.
Objects with measured fluxes lower than $0.1\mu$Jy have
been drawn with measured flux equal to $0.1\mu$Jy.  Error
bars for two representative galaxies are dotted.
The large circles with solid error bars show the mean
observed flux ($\pm$ standard deviation of the mean) for
galaxies in different bins of predicted flux.  The
galaxy with $f_{\rm pred}\sim 7$, $f_{\rm obs}\sim 200$
likely has a significant AGN contribution to its 20cm
flux; it was the only outlier rejected before calculating
the mean fluxes
}
\label{fnlband}
\end{figure}

An additional cause for concern is the fact that a red spectral
slope $\beta$ does not appear to be the most distinctive
characteristic of the high-redshift galaxies known
to have large dust luminosities.  For example, these are
the differences in mean color between $z\sim 1$ galaxies
in the HDF that are and are not ISO $15\mu$m sources:
$\Delta(U_n-G)=0.11$, $\Delta(G-{\cal R})=0.45$,
$\Delta({\cal R}-I)=0.17$.  The SEDs of the dusty and luminous
ISO sources differ from those of fainter and less obscured
galaxies primarily in the
range $2500\simlt\lambda_{\rm rest}\simlt 3500$\AA, not
$\lambda_{\rm rest}\simlt 2000$\AA\ where $\beta$ is measured.
This does not show
that MHC's prescription for dust correction
is wrong, only that it can be improved, but it
adds to concerns that their prescription
may not perform as well at high redshift
as in the local universe.  

Even if the $\beta$/far-IR correlation fails on some objects,
however, there is reason to hope that it may perform reasonably
well on an ensemble of high redshift galaxies.
For example, the correlation implies a mean dust obscuration for
$z\sim 3$ Lyman-break galaxies of 
$\langle L_{\rm bol,dust}/L_{\rm UV}\rangle\simeq 8$,
a plausible value that is midway between the dust obscuration
observed in local spirals ($\langle L_{\rm bol,dust}/L_{\rm UV}\rangle\simeq 5$)
and local UV-selected starbursts
($\langle L_{\rm bol,dust}/L_{\rm UV}\rangle\simeq 15$)\cite{as}
and that is consistent with the mean obscuration
$\langle L_{\rm bol,dust}/L_{\rm UV}\rangle\simeq 13\pm 6$
derived from sub-mm observations of Lyman-break galaxies
\cite{cet01}.  Moreover many of the major results to emerge
from recent analyses of $850\mu$m data -- 
the brightness of $850\mu$m background, the domination of
the $850\mu$m background by $\sim 1$mJy sources,
the $\sim$ thousand-fold increase in number density
of ULIRGs from $z\sim 0$ to $z\simgt 3$ -- could have
been predicted from UV observations alone by applying
the $\beta$/far-IR correlation to UV-selected high-redshift
galaxy populations\cite{as}.  Applying MHC's $\beta$/far-IR correlation
to high-redshift galaxy populations at least leads to results
that are reasonable, and so it seems sensible to continue using
it cautiously until further long-wavelength observations
can validate it or suggest better alternatives.

\section{Conclusions}
This conference concluded with a debate on the best way to
explore high-redshift star formation.  The best way?  Understanding the
history of galaxy formation is a large and difficult task,
a shared enterprise that will benefit from observations
at many wavelengths.  I am not sure how much we stand
to gain from arguing whose contributions are the most
important.

Nevertheless I would have felt derelict in my assigned role as
defender of the UV if I had not attacked one statement
that was repeated several times at the meeting.  This was the
claim that UV and IR surveys detect ``orthogonal'' populations
of galaxies, that a significant fraction of stars
form in a ``hidden'' population of dusty galaxies that is not detected
in UV surveys.  I am not aware of any evidence supporting
this claim.  Where is the dusty hidden population
in the local universe?  A luminosity-limited UV survey 
deep enough to detect most spirals and blue starbursts would
detect most LIRGs and ULIRGs as well (cf. Fig. 3).  Where is
it at $z\sim 1$?  Existing UV-selected
surveys easily detect most if not all of the $z\sim 1$ $15\mu$m sources
detected by ISO.  Where at higher redshift?  A UV-selected
survey would need to reach ${\cal R}\sim 26.5$ to detect
the majority of the $\sim 1800$\AA\ luminosity density
at $z\sim 3$\cite{set99}, and at this magnitude limit
most if not all of the $850\mu$m sources in the samples
of \cite{bcr}, \cite{iet}\footnote{including correction
for magnification by gravitational lensing}, and \cite{crlwb}
would be detected.  It is easy to construct a ``Madau diagram''
in which galaxies detected at wavelength $x$ contribute
far more to the comoving star-formation density
than galaxies detected at wavelength $y$.
Data will confess to anything if you torture them enough.
The proper approach is not to compare Madau diagrams -- not
to see who benefited most from the grotesquely compounded
uncertainties -- but to turn to the galaxies themselves
and measure their luminosities at both wavelengths.
When this is done (as we have seen) the answer is unequivocal:
any UV survey at a given redshift $z\simlt 3$ deep
enough to detect majority of the UV luminosity density will
detect the majority of IR-selected galaxies as well.

But there is one sense in which most star formation {\it is} hidden
from UV surveys:  only a small fraction of the energy emitted
by massive stars emerges from typical galaxies in the UV.
The majority is absorbed by dust and reradiated in the far-IR.
Unless there is some way to estimate the total luminosity
of a galaxy's massive stars from the (often) trace amount 
that is detected in the UV, 
it will be impossible to estimate star-formation rates
for the detected galaxies and the usefulness of UV surveys
will be diminished.
Sect. 3 discussed one method for estimating far-IR fluxes
from far-UV fluxes that is known to work in the local universe.
As we saw, this method appears to work within a factor
of $\sim 3$ at high redshift, but it is unclear if it works much
better.  The available data inspire more hope than confidence.

The strength of UV surveys is that they detect large numbers
of high-redshift galaxies, even ones that are intrinsically
very faint, in large and representative comoving volumes.
The weakness is that star-formation rates are difficult
to estimate for the detected galaxies.  
IR surveys complement them perfectly:  star-formation
rates can be estimated with reasonable confidence, but only
small regions of the sky can be surveyed and only the most luminous 
sources can be detected.  Harnessing the strengths of both types
of surveys would lead to a rapid advance in our understanding
of galaxy formation at high redshift.
One obvious strategy would be to use IR observations of some UV-selected
high-redshift galaxies to attempt to derive (or validate) a method
for estimating dust luminosities from UV observations alone,
a method that can
then be applied to the large numbers of UV galaxies which
have not been (and often cannot be) detected in the IR.
This sort of multiwavelength cooperation will take us much farther than
arguments about which wavelength is superior --
despite the best efforts of wavelength partisans like
the author of this screed.

\bigskip

I would like to thank the organizers for support at this enjoyable
meeting.  Special thanks are due to my collaborators
S. Chapman, M. Dickinson, M. Giavalisco, M. Pettini, A. Shapley, and 
C. Steidel for their contributions to this work.

%

\end{document}